\documentclass[sigconf]{acmart}

\usepackage{xcolor}  %
\usepackage{graphicx}
\usepackage{url}
\usepackage{hyperref}

\AtBeginDocument{%
  }

\setcopyright{acmlicensed}
\copyrightyear{2026}
\acmYear{2026}
\setcopyright{cc}
\setcctype{by-nc-nd}
\acmConference[SIGCSE Virtual 2026]{Proceedings of the 2nd ACM Virtual Global Computing Education Conference V.1}{November 12--15, 2026}{Virtual Event, USA}
\acmBooktitle{Proceedings of the 2nd ACM Virtual Global Computing Education Conference V.1 (SIGCSE Virtual 2026), November 12--15, 2026, Virtual Event, USA}
\acmDOI{10.1145/3795867.3830962}
\acmISBN{979-8-4007-2506-7/2026/11}

\begin{document}



\title[A Tool to Map AI Programs in the U.S]{A Tool to Map AI Programs in the U.S.: A Snapshot from April 2026 and an Analysis of Requirements for AI Majors and Minors}

\author{Felix Muzny}
\email{f.muzny@northeastern.edu}
\orcid{0000-0002-0038-4222}
\affiliation{%
\institution{Center for Inclusive Computing}
  \institution{Northeastern University}
  \streetaddress{177 Huntington Ave}
  \city{Boston}
  \state{Massachusetts}
  \country{USA}
  \postcode{02115}
}

\author{Carolyn Jones}
\email{car.jones@northeastern.edu}
\orcid{0009-0001-0050-508X}
\affiliation{%
\institution{Center for Inclusive Computing}
  \institution{Northeastern University}
  \streetaddress{177 Huntington Ave}
  \city{Boston}
  \state{Massachusetts}
  \country{USA}
  \postcode{02115}
}

\author{Carter Ithier}
\email{c.ithier@northeastern.edu}
\orcid{0009-0003-1949-5437}
\affiliation{%
\institution{Center for Inclusive Computing}
  \institution{Northeastern University}
  \streetaddress{177 Huntington Ave}
  \city{Boston}
  \state{Massachusetts}
  \country{USA}
  \postcode{02115}
}
\author{Hasnain Sikora}
\email{sikora.h@northeastern.edu}
\orcid{0009-0004-4949-4053}
\affiliation{%
  \institution{Center for Inclusive Computing}
  \institution{Northeastern University}
  \streetaddress{177 Huntington Ave}
  \city{Boston}
  \state{Massachusetts}
  \country{USA}
  \postcode{02115}
}

\author{Hrutika Harshadbhai Patel}
\email{patel.hru@northeastern.edu}
\orcid{0009-0001-1466-7178}
\affiliation{%
  \institution{Center for Inclusive Computing}
  \institution{Northeastern University}
  \streetaddress{177 Huntington Ave}
  \city{Boston}
  \state{Massachusetts}
  \country{USA}
  \postcode{02115}
}

\author{Carla E.\ Brodley}
\email{c.brodley@northeastern.edu}
\orcid{0009-0008-2134-6285}
\affiliation{%
  \institution{Center for Inclusive Computing}
  \institution{Northeastern University}
  \streetaddress{177 Huntington Ave}
  \city{Boston}
  \state{Massachusetts}
  \country{USA}
  \postcode{02115}
}

\renewcommand{\shortauthors}{Felix Muzny et al.}

\begin{abstract}
    In this work, we locate and analyze existing undergraduate Artificial Intelligence (AI) programs in the United States in Spring 2026, creating a historic record at a time of great change in this area.
    To create this record, we developed a tool to detect, scrape, and display data from 361 undergraduate AI programs--majors, minors, concentrations, and certificates--at 4-year universities.
    Our tool, available at \href{https://cicmap.ai}{https://cicmap.ai}, searched 563 institutions to locate these programs, a sample that represents 87\% of all undergraduate Computer Science (CS) graduates in the U.S in 2025.
    This tool allows prospective students, guidance counselors, administrators, and faculty to easily access AI program requirements and is designed to continually update as new programs emerge.
    To the best of our knowledge, this survey represents the most comprehensive snapshot of the state of AI programs in the U.S. to date.
    With this work we offer three important contributions: 1) a record of AI programs in the U.S. at a time of great upheaval; 2) a tool to explore AI programs and their requirements; and 3) an analysis of the courses required for 66 AI majors and 87 AI minors.
    Our analysis of majors and minors shows great variability in the size and the requirements of these degrees, but we note two takeaways.
    First, not all majors require a general AI course, but if they don't, they do require a Machine Learning (ML) course. Second, more than a third of majors require an Ethics in AI course but only 24\% of minors do.

\end{abstract}

\ccsdesc[500]{Social and professional topics~Computer science education}
\ccsdesc[500]{Social and professional topics~Model curricula}

\keywords{Education, Artificial Intelligence, Data Scraping, AI Majors, AI Minors}

\maketitle

\section{Introduction}

While there is a sense that AI programs in higher education have exploded in recent years, we lack a comprehensive snapshot of where these programs are actually located, either globally or in the U.S. specifically.
The disconnect between the demand for these programs and the resources available to students and administrators for locating them will only increase until this information gap has been addressed.
To this end, we developed a tool to bridge this gap.  In this paper we present our tool and provide a snapshot of the state of AI programs in the U.S. as of April 2026.  The results of running our tool are publicly accessible. Due to the level of automation, these results will continue to allow access to up-to-date information about the location and contents of these programs.

When considering AI courses and programs, these vary in style from AI literacy to technical AI skills.
In terms of AI literacy, there are calls for all U.S. universities to ensure their graduates have these skills  \cite{forbesFluencySupport, upceaLiteracy}.
Indeed, some sources state that AI literacy is a requirement for graduates to stay competitive on the job market \cite{forbesAIEmployeerNeed}.
In the U.S., states are currently writing and implementing policy mandating that AI literacy be a part of K-12 education \cite{niaie2025stateai, georgia_sb179_2025, mississippi_sb2294_2026, nj_a4352_2026, nj_s2862_2026}.
For the subset of students interested in becoming technical experts in the development of AI models, AI literacy only scratches the surface.
These students need the skills to understand the math, data management, resource allocation, and other technical underpinnings from core CS curricula to become experts in AI model development.

Programs that offer students credentials affirming their specialized technical AI skills include majors, minors, concentrations,\footnote{We use the term ``concentration'' as the general term for specializations, tracks, and concentrations within a degree.} and certificates in AI, but, these programs can be difficult to find.
While a search engine or generative AI tool may be able to provide a cursory list of institutions with these programs, it is difficult for a single person to understand the overall landscape of AI programs, their locations, and their content from these searches.
Zooming in on program content, top-level pages surfaced by search tools are often focused on marketing rather than the actual program requirements, which are often buried in complex menus and course catalogs.
This makes finding and understanding AI-specific programs especially difficult for prospective and transfer students who may be first-generation or lack guidance resources at their high schools and community colleges.

With this work, we aim to bridge these critical gaps by developing tools to automatically detect and map AI programs within computing departments and colleges in the U.S.
The tools we develop attempt to include every non-profit 4-year institution, regardless of size, whether it is public or private, how research-active the faculty are, and the university/department rank.

We have successfully scraped information from 563 institutions, which collectively produced 87\% of U.S. CS graduates in 2025 as reported by the Integrated Post Secondary Education Data System (IPEDS) \cite{ipeds}.\footnote{Defined as Classification of Instructional Programs (CIP) codes 11.0101 and 11.0701.}
Among these institutions, we found a total of 73 AI majors and 89 AI minors and we know this number is growing.
Of these programs, 66 AI majors and 87 AI minors had complete and publicly accessible requirements pages as of April 2026.

To increase ease-of-access to the information we find, we created an interactive map interface, allowing users to filter by various features such as program type, state, and the number of BS/BA students the CS department graduates per year. 
This interface includes links to the websites that describe each program's course requirements so an explorer can easily find a listing of the actual content of the program.
Our map is designed to dynamically update, correcting for reported errors and detecting newly-launched programs each time we re-run our scraping tools.\footnote{Due to constraints on compute available and amount of time to process each university, we are currently unable to sustain a faster full refresh rate than every three months.}

By tracking when these programs launch, not only does our tool provide an easy-to-access resource for students, administrators, and researchers, it allows us to track the emergence of these programs in the U.S. over the next several years.

\begin{figure*}[h!]
    \centering
    \includegraphics[width=0.85\textwidth]{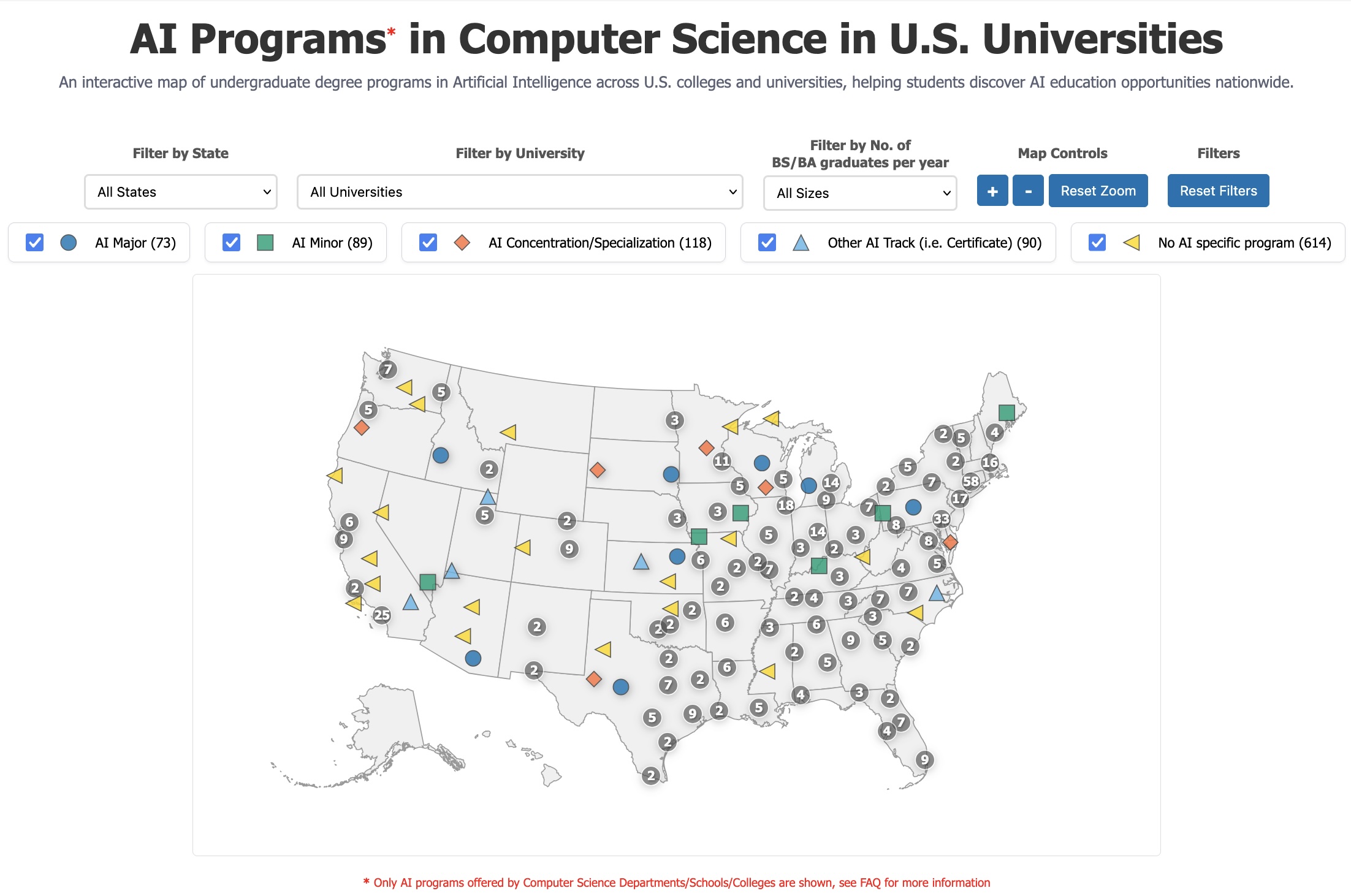}  
    \caption{The interactive map generated by our tool, visualizing 975 AI and CS programs at 563 institutions.}
    \label{fig:map_desktop}
\end{figure*}

\section{Background}

To the best of our knowledge, this is the most comprehensive review of AI programs both in the U.S. and in general to date. 
There have been a few worthy starts at characterizing the AI program landscape in recent years.
In a 2024 work in progress, He et al. describe AI majors at eight top-tier Chinese and American universities, identifying that 21 - 41\% of required courses within AI majors at these universities are AI courses \cite{he2024wip}.
A 2025 report from Stanford University identified 19 U.S. universities offering Bachelor's degrees in AI as of 2023 which they determined from IPEDS graduation data, and reported that 104 students graduated with these degrees \cite{maslej2025artificial}, a number they expect to rise. The Computing Research Association has convened dozens of educators through the Level-UP AI initiative to understand the state of AI education across North America and has published a report on their findings \cite{Level-up-AI}, but has not yet provided a summary of the education opportunities available.  
In the 2026 version of the Stanford report, the authors note increases in numbers of post-secondary students earning AI-related\footnote{These are split into two categories: 1) AI software, which includes majors such as Artificial Intelligence, Computer Programming, and Computational and Applied Mathematics; and 2) AI hardware, which includes majors such as Electrical and
Electronics Engineering, Condensed Matter and Materials Physics, and Industrial Engineering.} degrees, with particularly large increases among Masters students \cite{aiindex2026}.

Turning to reviews of individual AI courses, a 2024 Master's thesis inspected a random sample of eighty highly-research active universities in the U.S. and Canada \cite{niousha2024artificial}.
This work counted the number of AI courses available within these universities by hand and identified pre-requisites for AI courses, comparing and contrasting U.S. versus Canadian institutions, finding that the most common number of AI courses for a university to have was three, with greater variation in Canada than in the U.S.

When it comes to CS curriculum characterizations as a whole, approximately every ten years an Association for Computing Machinery (ACM) Joint Task Force issues curriculum guidelines for what a CS degree should contain.
This is an attempt to coalesce opinions about the structure of CS programs from a broad group of academics, representing a wide swath of universities in the U.S. though we could not find documentation of any universities applying the guidelines to their curriculum as a whole.

In the CS2023 ACM curricular guidelines, the task force identified AI as one of their 17 \textit{knowledge areas} and recommended 4.4\% of the core hours (12 of 270) in a CS degree should be devoted to this topic \cite{acm2023curricula}.
This is a change from 2013, when the committee did not allocate any of the CS core hours to AI \cite{acm2013curricula}.
The \textit{knowledge units} within AI were defined to be: fundamental issues, search, fundamental knowledge representation and reasoning, machine learning (ML), applications and societal impact, probabilistic representation and reasoning, planning, logical representation and reasoning, agents and cognitive systems, natural language processing (NLP), robotics, and computer vision.
Among these knowledge units, the only units denoted as recommended for the CS core were fundamental issues, search, fundamental knowledge representation and reasoning, ML, and applications and societal impact. 
The guidelines present learning outcomes, such as designing the state space representation for a puzzle, alongside suggestions for groups of knowledge units and sub-topics for AI, ML, Robotics, and Data Science (DS) courses.
Though this committee aims to receive input from many universities, it's goal is not to provide a snapshot of what universities are actually doing, but, rather, to provide recommendations for what they should do.

A 2026 study from Xai et al. worked to analyze alignment of topics within AI courses with these CS2023 ACM guidelines, but the analysis does not move beyond word counts and clustering courses based on name similarity \cite{xai2026comparativeanalysisai}. 
According to these metrics, they find that existing courses align well with CS2023 guidelines.

Finally, we note criticism of the CS2023 ACM guidelines. 
One such voice comes from a 2025 perspective from Holland-Minkley et al. that is informed by liberal arts CS programs \cite{holland2025liberalartsacm}.
Holland-Minkley et al.'s overall argument is that no model curriculum can capture the diversity of such CS programs, and, instead, propose a process for applying guidelines to support a diversity of collegiate CS education paradigms.
They note how institutional missions can align with departmental visions and identity statements to identify intended institutional emphasis and thus guide the CS curriculum appropriately within the institution's context. 
While the focus of their work was not to measure alignment between CS2023 and the curricula in AI majors, it is important context for our analysis.

\begin{figure*}[h!]
    \centering
    \includegraphics[width=0.75\textwidth]{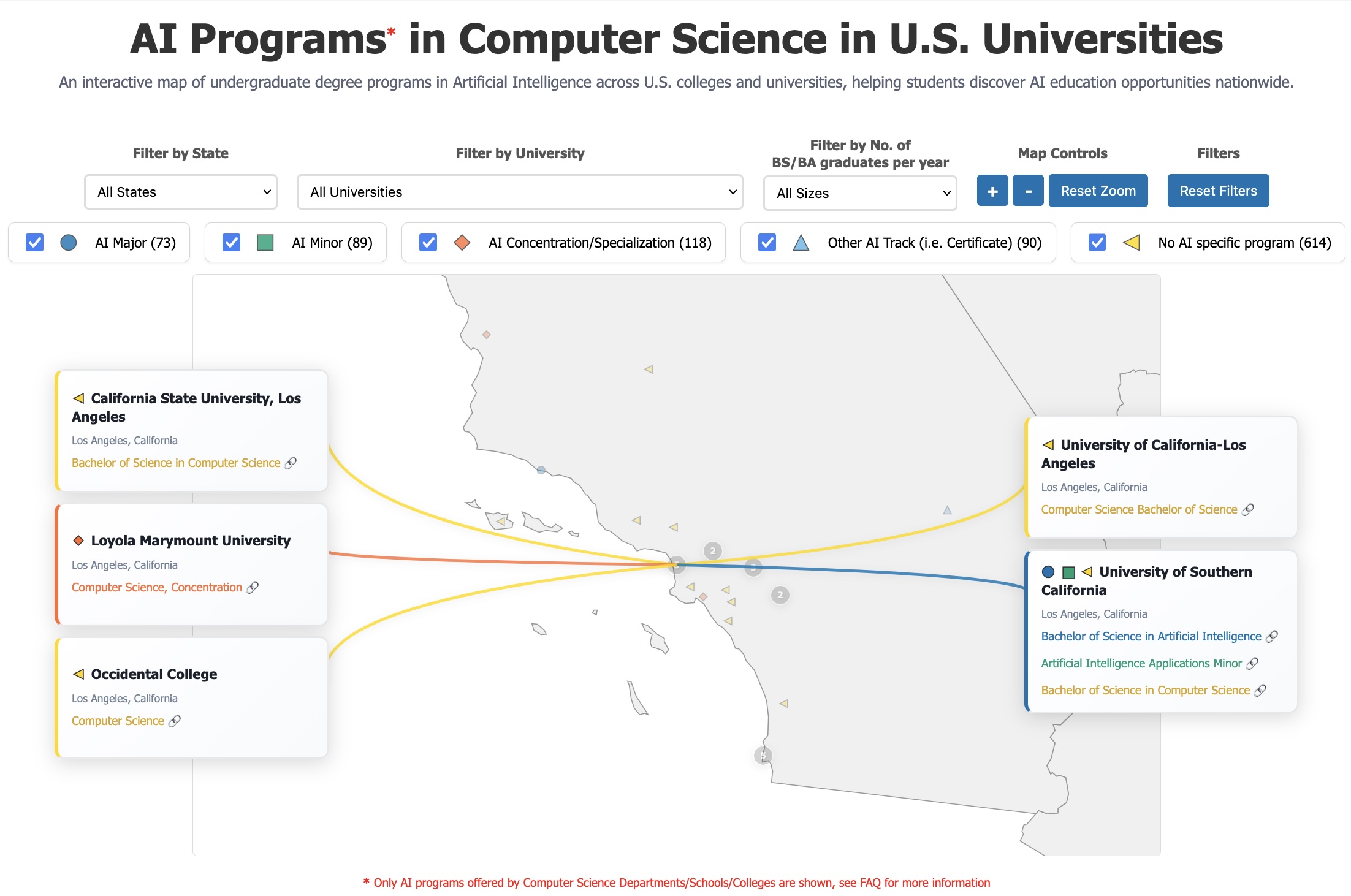}  
    \caption{When clusters are clicked, the map zooms in and annotations appear with links directly to program requirements.}
    \label{fig:map_socal}
\end{figure*}

\section{A Tool to Map AI Programs}

Figure ~\ref{fig:map_desktop} shows the map of AI and CS programs our tool generates with each school represented by a symbol denoting the program type.
To create this resource, we developed a pipeline of scraping tools capable of scraping large amounts of university, program, and curriculum data from the web. 
The first scraping tool uses Google and Exa search APIs to find candidate university websites, given a list of target universities.
These websites are then filtered using DeepSeek, an open-source large language model (LLM), to identify departmental websites for each university.
From these websites, we are able to search for both traditional CS and for AI degree pages.

Next, we use a human-in-the-loop to verify the correctness of the identified websites before using them as input for the second tool. 
Using a human-in-the-loop was critical because even with multiple different search APIs, both traditional and those powered by LLMs, we had a failure rate of around 33\% for the links initially identified.
Humans were able to mark identified links according to correctness and find the links the automatic tools were unable to locate.
While this does involve manual labor, we anticipate that because program website addresses rarely change, the initial investment of time means that future iterations will have few updates to verify.

Our second scraping tool uses DeepSeek to identify specific program type (e.g., major, minor, etc.) and program name as well as to gather data about the curriculum itself, including required courses for the program, to be used in future research. 

We conducted validation throughout the data collection process. 
During development, we manually labeled a sample of 113 programs with program type labels; our tool achieved 100\% accuracy in program type identification.  
Because of the critical nature of the correctness of this resource, we hand verified the labels of all other programs included in our map before releasing it to the general public at \textit{\href{https://cicmap.ai}{https://cicmap.ai}}. 

We use the data gathered by these tools to power an interactive map, shown in Figures \ref{fig:map_desktop} and \ref{fig:map_socal}, that allows users to explore the U.S. geographically, filter results by various attributes, and view information about specific programs when selected.
Combining the data we scraped with information from IPEDS allows us to add filtering functions by state, university, degree type, and the number of CS graduates per year to our tool.
In Figure \ref{fig:map_desktop}, the gray circles with numbers represent geographical concentrations of institutions, with each cluster labeled to indicate the number of institutions it contains. This clustering approach serves two key purposes: 1) it prevents visual overcrowding, and 2) it enables users to quickly identify densely populated regional hubs. Users can click on any cluster to zoom into that region, revealing individual institution markers with detailed program information (see Figure ~\ref{fig:map_socal}).

\section{Findings}
\label{sec:findings}

As stated earlier, we have scraped data from 563 institutions thus far, representing 87\% of all 2025 U.S. CS graduates.
Those institutions we did not include are those we were unable to scrape data from because of website formatting issues.\footnote{This occurs when newly released programs are not web-discoverable, when links on departmental websites are dead or missing, and when convoluted formats cause errors when converting websites to the formats we use internally such as pdf and markdown.}
We found that 249 out of the 563 (44\%) institutions we scraped had one or more AI programs.
We expect this percentage to rise dramatically as universities develop more AI programs to meet the demands of industry and academia.

In terms of program type, concentrations were the most common AI program detected, encompassing 32.7\% of the AI programs found (118 out of 361).
We hypothesize that the concentration format allows administrations and CS departments to respond rapidly to the growing demand for AI education, as concentrations can be developed within existing degree structures and may face fewer administrative barriers, whereas a new AI degree requires a multi-stage approval process which can take 1-2 years.

We have already witnessed numerous programs rise since we started this project, with the number captured by our tool climbing from 62 at the end of January 2026 to 73 in April 2026.
Our tool is able to catch newly developed programs because we re-scrape the data roughly once every three months and allow corrections to be submitted by users off-cycle.
Although we found that over half of institutions do not offer an AI-specific program, that does not mean they do not teach AI content. Many offer elective courses in AI topics such as Foundations of AI, ML, NLP, Computer Vision, etc. 
We leave analysis and detection of these courses to future research.

To characterize the programs we located by size and geographic distribution, we use IPEDS location and graduation data.
We categorize the \textit{size} of institutions based on \textit{number of graduates} under Category of Instructional Programs (CIP) codes 11.0101 and 11.0701. These codes map to general CS degrees.
A small institution has fewer than 25 graduates per year, medium with 100 or fewer, and as large with more than 100 graduates.\footnote{Institutions categorized as "Unknown" are outlying programs whose CS or technical degrees do not fall under CIP codes 11.0101 and 11.0701 and thus lack IPEDS completion data for size classification.}
14.4\% (17) of small institutions scraped (118) have AI majors, 10.2\% (12) have minors, 14.4\% (17) have concentrations, and 16.1\% (19) have other AI programs. Note that an institution can have more than one AI program and are counted in each category for which they have a program. 
We scraped 279 medium institutions, of which 6.8\% (19) have majors, 14.0\% (39) have minors, 20.4\% (57) have concentrations and 10.0\% (28) have other AI programs.
We scraped 169 large institutions, of which of which 18.3\% (31) have majors, 17.2\% (29) have minors, 21.9\% (29) have concentrations and 16.0\% (27) have other AI programs.  We were unable to analyze the programs at 3 universities.  Finally, 52.5\% (62) of small, 60.6\% (169) of medium, 47.7\% (81) of large institutions have no AI programs.

Taking a brief geographic snapshot, we see the highest number of AI programs in the following regions of the U.S.: Southeast,\footnote{AL, AR, FL, GA, KY, LA, MS, NC, SC, TN, VA, and WV} Mid East,\footnote{DE, DC, MD, NJ, NY, and PA} and the Great Lakes.\footnote{IL, IN, MI, OH, and WI} 
However, the number of AI programs also correlates with the number of universities in the region, with less populous regions having fewer universities.


\begin{figure*}[h!]
    \centering
    \includegraphics[width=0.8\linewidth]{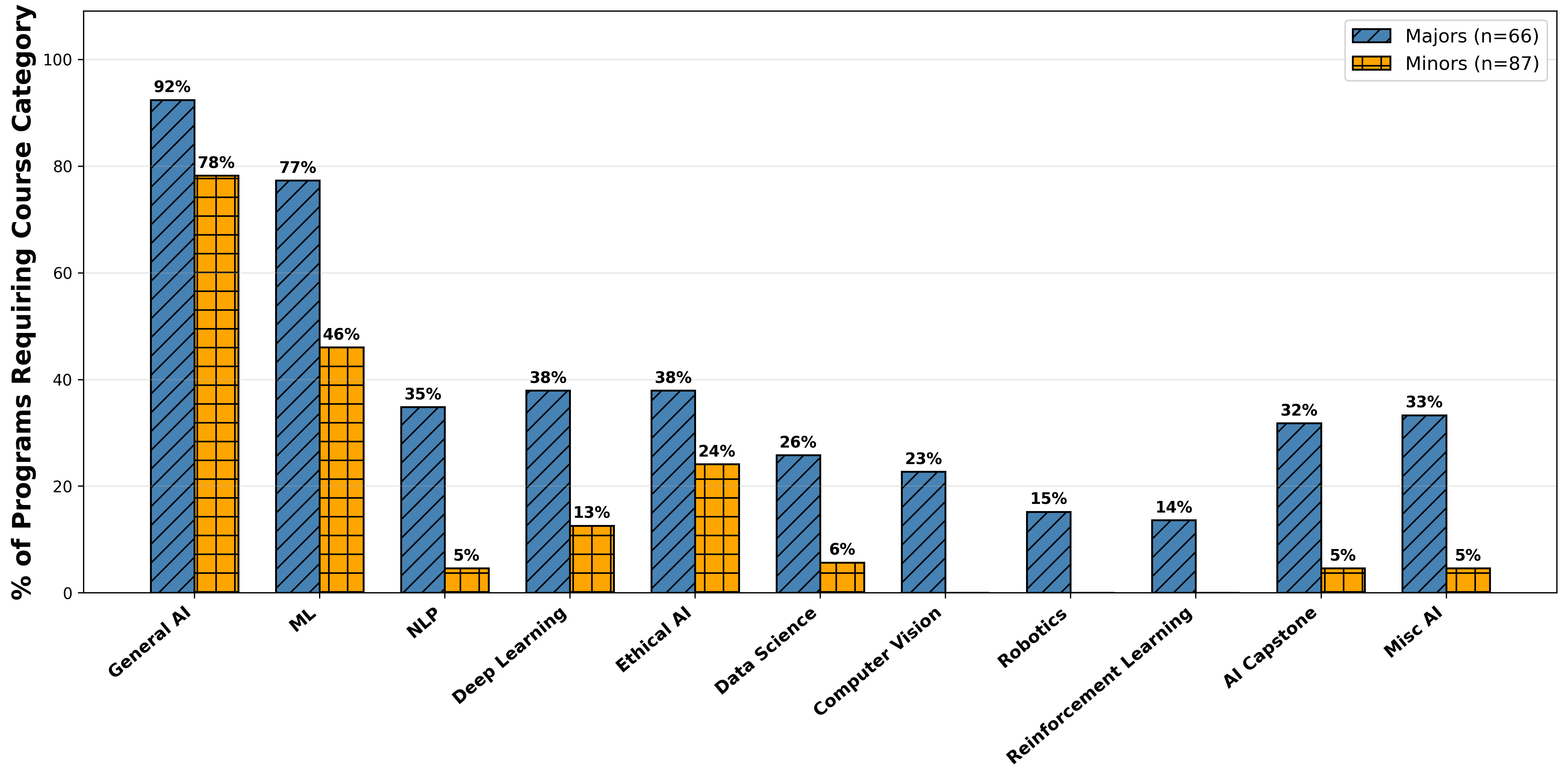}
    \caption{Requied AI courses for 66 AI majors and 87 AI minors.}
    \label{fig:major_course}
    
\end{figure*}

\subsection{What do AI Majors Require?}

For the 66 AI majors we are able to analyze, we calculated the following metrics, all in terms of credit counts: 1) total to graduate, 2) major requirements,\footnote{Including required supplemental math and science courses, and required core major courses.} 3) required AI credits,\footnote{Any core course teaching AI-related technical concepts or ethics of AI.}
4) CS credit requirements,\footnote{Not including AI related credits.} and 5) major electives. 
While a typical Bachelor's degree in the U.S. consists of 120 credits, not all do, so credits were normalized to 120.  
We report these statistics in Table \ref{tab:ai_major_credits}.

The data shows significant variability in the breakdown of credits within AI majors. 
Proportionally speaking, an AI major may include anywhere from 30\% to 89\% of all credits necessary for Bachelor's degree completion, reflecting the wide variability in how U.S. universities distribute requirements between major credits and general requirements.
When we consider the breakdown of credits within the degree into AI-specific credits and CS-specific credits (non-inclusive of AI courses or non-CS courses), these programs also show great variability.

AI-specific credits are the credits from courses that teach either technical aspects of AI or the ethics of AI. 
These courses can be categorized with the following labels: introductory AI (before either data structures, or the third introductory CS course if no data structures), AI (after data structures, as in a typical junior- or senior-level AI course), ML, ML/AI combined, NLP, Computer Vision, Deep Learning/Neural Networks, Reinforcement Learning, Robotics, DS, Responsible/Ethical AI, Capstone AI, and Miscellaneous AI. 
We make these categorizations based on catalog descriptions of each course to correct for differences in naming conventions.
We further break AI-specific credits into two subcategories: 1) those from specifically required courses (e.g., ``all AI majors must take \textit{Foundations of AI}'') and 2) those from elective courses (e.g., ``take 9 credits from the \textit{Applications of AI} list'').
We call credits from specifically required AI courses \textit{required AI credits} and place those from elective AI courses into the \textit{elective AI credits} category, as shown in Table \ref{tab:ai_major_credits}.

\begin{table}[H]
\small
\centering
\begin{tabular}{lcccc}
\toprule
\textbf{Metric} & \textbf{Mean} & \textbf{Std Dev} & \textbf{Min} & \textbf{Max}  \\
\midrule
Required Major Credits & 65.5 & 16.4 & 36.0 & 106.7 \\ 
\% of total Credits to graduate & 54.6\% & 13.7\% & 30.0\% & 88.9\% \\ \hline
Required AI Credits & 18.1 & 8.2 & 3 & 42  \\
\% of Major that is Required AI & 26.2\% & 12.4\% & 4.8\% & 69.2\%  \\ \hline
Required Elective AI Credits & 9.9 & 7.2 & 0.0 & 27.0 \\
\% of Major that is AI Electives & 9.9\%  & 7.2\%& 0\% & 53.8\% \\ \hline 
Required CS, non-AI Credits & 19.3 & 10.0 & 3 & 39.0  \\
\% of Major that is CS (i.e., non-AI) & 29.4\% & 14.0\% & 7.1\% & 76.2\% \\ 

\bottomrule
\normalsize
\end{tabular}

\caption{Required credits for 66 AI majors, separated by required AI and required CS, non-AI, courses. Credits in the mean column do not sum to 120 because credits from non-CS, non-AI courses are not included.}
\label{tab:ai_major_credits}
\end{table}

Notably, there is little consistency in proportionality for the breakdown of credits within AI majors.
Table \ref{tab:ai_major_credits} shows that for required AI credits, there is a maximum of 42 and a minimum of 3 credits.
Programs at the lower end of the required AI credit spectrum often reflect two things: 1) a greater proportion of elective AI credits and/or 2) an interdisciplinary approach that emphasizes data science, data analytics, or statistical learning alongside AI techniques. 
Another factor is that institutions distribute the bulk of AI credits differently, some focusing on computational basics in the required core (more required CS, non-AI credits) before moving on to AI-related material in the electives category while others include more specific AI courses in the required set. 

Employers of students with AI majors with more required AI credits listed can expect them to have a more consistent background while students coming from programs with more elective AI credits have greater opportunity to specialize in a particular aspect of AI.

Next, we investigate the specific courses that are required in AI majors across universities.
Shown in Figure ~\ref{fig:major_course}, we found that 92\% of all AI majors require some form of general AI course.
This category includes introductory AI (before data structures), AI (after data structures), and ML/AI courses.

It may seem surprising that not all AI majors require a general AI course--we find that those not requiring a general AI course do require an ML course.
One example is a program that requires an ML foundation course but offers general AI courses like``Intro Artificial Intelligence'' as electives.
ML was the second most required class, with 77\% of the 66 majors analyzed requiring at least one.   
Deep Learning, Responsible/Ethical AI, and NLP were the next most common individual courses with slightly more than a third of all AI majors requiring each one.

Inspecting combinations of required courses, the largest overlapping set seen in at least five programs is \{Computer Vision, Ethical AI, General AI, ML, NLP\} with five programs (7.6\%) requiring all five of these courses--though this may not depict the complete set of required courses in these programs.
The most common combinations of four courses are \{I Capstone, Ethical AI, General AI, ML\} and \{Deep Learning, General AI, ML, NLP\}, both with nine programs (13.6\%) requiring them.
Popular three-course sets are shown in Table \ref{tab:major_course_combinations}.
46 programs (69.7\%) require at least both (General AI, ML) and 24 (36.4\%) require (Ethical AI, General AI).
\footnote{The complete list of combinations and their counts is available at \href{https://github.com/muzny/tool-to-map-ai-programs}{https://github.com/muzny/tool-to-map-ai-programs}.}
Overall, this data reflects the relative flexibility or rigidity of different AI majors.

\begin{table}[h]
\centering
\begin{tabular}{lrr}
\hline
\textbf{Course Combination AI Majors} & \textbf{Count} & \textbf{\%} \\
\hline
Ethical AI, General AI, ML       & 21 & 31.8 \\
Deep Learning, General AI, ML    & 17 & 25.8 \\
General AI, ML, NLP               & 17 & 25.8 \\
AI Capstone, General AI, ML      & 16 & 24.2 \\
General AI, ML, Misc AI          & 14 & 21.2 \\
AI Capstone, Ethical AI, General AI & 11 & 16.7 \\
Computer Vision, General AI, ML  & 11 & 16.7 \\
Data Science, General AI, ML     & 11 & 16.7 \\
\hline
\end{tabular}
\caption{Number and percent of 66 AI majors requiring each three course combination for combinations appearing in more than 10 programs. Note a single major may be counted in multiple rows.}
\label{tab:major_course_combinations}
\end{table}


\subsection{What do AI Minors Require?}

We conducted the same analysis for AI minors, counting total minor credits, required AI course credits, and required AI elective credits.
Table \ref{tab:ai_minor_credits} shows the statistics the 87
minors we located.
The total credit counts for AI minors were more standard than for the AI majors, resulting in smaller standard deviations. On average, a third of an AI minor consists of required AI courses, excluding other electives students can take. The rest of the credits in the minor are taken by computing classes, math classes, and other support/prerequisite courses.
We note that minors vary in the degree to which they explicitly state prerequisites that are implicitly required, which may make them more difficult to complete than first appears \cite{lionelle2026minors}.

\begin{table}[h]
\centering
\begin{tabular}{lcc}
\toprule
\textbf{Metric} & \textbf{Mean} & \textbf{Std Dev} \\
\midrule
Required Minor Credits & 19.1 & 4.7 \\
\% of total credits to graduate & 15.9\% & 3.2\% \\ \hline 
Required AI Credits & 6.4 & 3.3 \\
\% of Minor that is Required AI & 34.8\% & 19\% \\ 
\hline
Required Elective AI Credits  & 5.5 & 4.0 \\
\% of Minor that is AI Electives & 30.4\% & 22.3\% \\ 
\bottomrule
\end{tabular}
\caption{Required credits for AI minors, separated by those from required AI courses and those from electives.}
\label{tab:ai_minor_credits}
\end{table}

Shown in Figure \ref{fig:major_course}, AI minors require general AI courses most frequently, with 78.2\% of all minors requiring these courses. 
A smaller fraction of minors (24.1\%) than of majors--where it was 37.9\%--require an AI-specific ethics course.

We find that most AI minors designate few specific courses to be taken.
The biggest difference with the AI majors is that the two individually most-required courses in AI minors, General AI and ML, are not \textit{both} required in any of the 87 minors studied. 
Only five programs require three specific courses, 19 more specify two courses, and 59 specify only one course.
Four programs do not name any specific courses, instead letting students choose from a selection to fulfill all minor requirements.

\section{Conclusions and Future Work}
\label{sec:conclusion}

Our interactive map allows for a comprehensive view of AI programs in the United States, mapping 975 AI and CS programs to the 563 institutions scraped.
These universities collectively produce 87\% of the U.S.'s CS graduates in 2025. 
The tool serves as a critical resource for students seeking AI education, administrators benchmarking their programs, departments creating new programs, and researchers investigating trends in AI.
To the best of our knowledge, there is currently no other tool that can provide such information.

An important topic we have not yet confronted is the overlap between DS and AI, both at the program level and at the course level.
To control the scope of this project, we have not yet expanded our tools to include DS programs, but believe that is a worthwhile and necessary next step.
Future work may ask what, if anything, differs between DS and AI programs.

Our findings reveal significant opportunity for growth in AI education in the U.S. 
In April 2026, over 40\% of institutions in the U.S. currently offer AI programs, a number we know is rising. 
As this landscape continues to evolve, we intend to expand the capabilities of our mapping tool.
In the future, we will not only continue to add data from universities as they launch new AI programs, we will also pay special attention to specific barriers students face in accessing these programs. 
For example, can students who are not ready to take Calculus I when they enter university complete these degrees in four years? 
To this end, we will expand this work to gather and analyze specific course data in more depth, enabling us to answer questions about the structures of typical AI programs.

\begin{acks}
This work was funded by the NSF (grant \#2533723), Pivotal, and Northeastern University. 
\end{acks}

\balance
\bibliographystyle{ACM-Reference-Format}
\bibliography{mybib}

\end{document}